%% file: main.tex
\documentclass[11pt,twoside]{scrartcl}
\KOMAoptions{paper=a4,fontsize=11pt,pagesize=pdftex,headinclude,BCOR=8.5mm,DIV=15}

\include{header}

\begin{document}

\include{titlepage}

\clearpage

%\tableofcontents

%\clearpage

\section*{Hintergrund}
\subsection*{Der Weg zum Mars}
\begin{figure}[!ht]
    \centering
    \includegraphics[width=10cm]{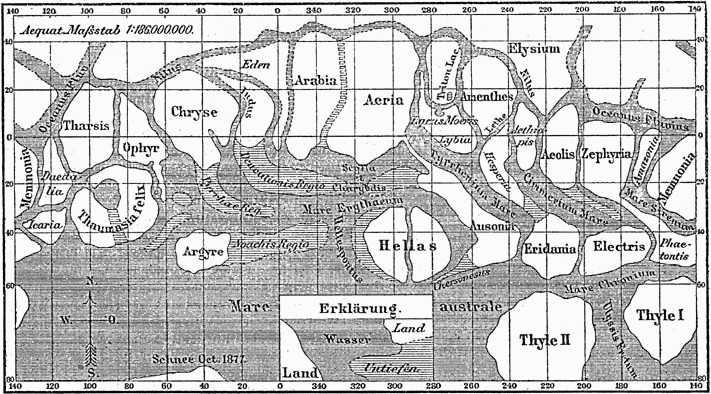}
    \caption{Karte der Marsoberfläche von 1877 nach Beobachtungen von Giovanni Schiaparelli (Quelle: Meyers Konversations-Lexikon, 1888, \url{https://commons.wikimedia.org/wiki/File:Karte_Mars_Schiaparelli_MKL1888.png}).}
    \label{fig:schiaparelli}
\end{figure}

Der Mars ist wohl der Planet im Sonnensystem, der die meisten Fantasien weckt. Spätestens seitdem Giovanni Schiaparelli 1877 auf dem Mars angeblich Kanäle entdeckt hatte \autocite[Abb.~\ref{fig:schiaparelli}][]{meyer_meyers_1885,schiaparelli_decouvertes_1882,schiaparelli_observations_1897}, gilt er als der Himmelskörper, der in puncto außerirdisches Leben am meisten fasziniert. Besondere Berühmtheit erlangte das Buch „War of Worlds (Krieg der Welten)“ von H. G. Wells aus dem Jahre 1898. Ein darauf basierendes Hörspiel erzeugte während seiner Radioausstrahlung 1938 Irritationen unter der Bevölkerung der USA, die die Sendung als einen Bericht missverstand \autocite[][]{chilton_war_2016}.

\begin{figure}[!ht]
    \centering
    \includegraphics[width=10.0cm]{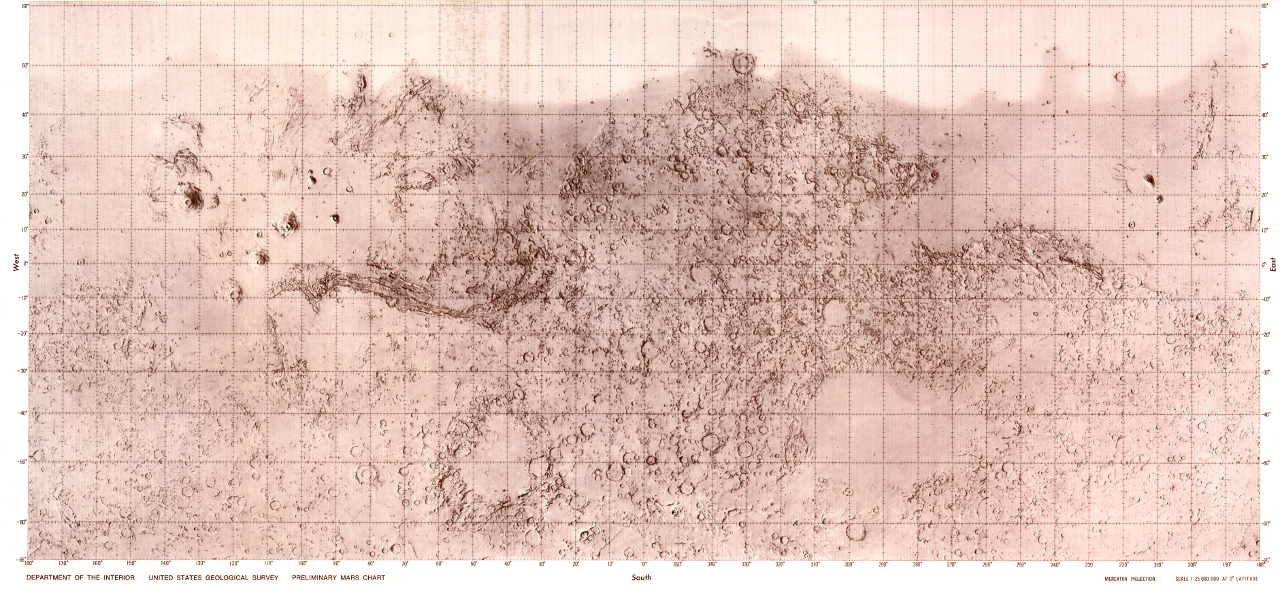}
    \caption{Karte der Marsoberfläche erstellt aus Fotografien der Marssonde Mariner~9 (Bild: USRA -- Universities Space and Research Association, Lunar and Planetary Catalog, \url{https://www.lpi.usra.edu/resources/mars_maps/mariner9/index.html}).}
    \label{fig:mariner9}
\end{figure}

Jenseits dieser fantastischen Erzählungen versuchen Wissenschaftler bereits seit dem Beginn der Raumfahrt, den Mars genauer zu erkunden. Die erste erfolgreiche Mission zum Mars war die Sonde Mariner~4, die 1965 auf ihrem Vorbeiflug die ersten Aufnahmen der Oberfläche aus unmittelbarer Nähe zur Erde funkte\footnote{\url{https://nssdc.gsfc.nasa.gov/nmc/spacecraft/display.action?id=1964-077A}}. Als erster künstlicher Marssatellit schwenkte Mariner~9 1971 in den Orbit ein\footnote{\url{https://nssdc.gsfc.nasa.gov/nmc/spacecraft/display.action?id=1971-051A}} und erstellte die erste detaillierte Marskarte (Abb.~\ref{fig:mariner9}).

Die ersten Landungen auf dem Mars gelangen 1976 mit \mbox{Viking 1\footnote{\url{https://nssdc.gsfc.nasa.gov/nmc/spacecraft/display.action?id=1975-075C}}} und Viking 2\footnote{\url{https://nssdc.gsfc.nasa.gov/nmc/spacecraft/display.action?id=1975-083C}}. Sie definierten bereits die grundlegenden Technologien für weiche Landungen auf dem Mars, die unter anderem aus Fallschirmen und Bremsraketen bestanden. Dass Flüge zum Mars generell eine große technologische Herausforderung sind, erkennt man daran, dass nur ca. 50\% aller Marsmissionen erfolgreich sind \autocite[][]{zundel_mars-mission_2016}.

\subsection*{Landung auf Erde und Mars}
Wenn Raumschiffe wie früher Apollo oder noch heute die Sojus-Kapseln auf der Erde landen, werden sie durch Fallschirme auf Geschwindigkeiten abgebremst, die eine nahezu weiche Landung ermöglichen. Der Aufprall der Apollo-Raumschiffe wurde durch eine Wasserung im Ozean abgemildert, während Düsen an den Sojus-Landemodulen kurz vor dem Aufsetzen zünden und die Restgeschwindigkeit auf akzeptable Werte reduzieren\footnote{\url{https://www.esa.int/kids/de/lernen/Leben_im_Weltraum/Astronauten/Achterbahnfahrt_zurueck_zur_Erde}}.

\begin{figure}[!ht]
    \centering
    \includegraphics[width=7cm]{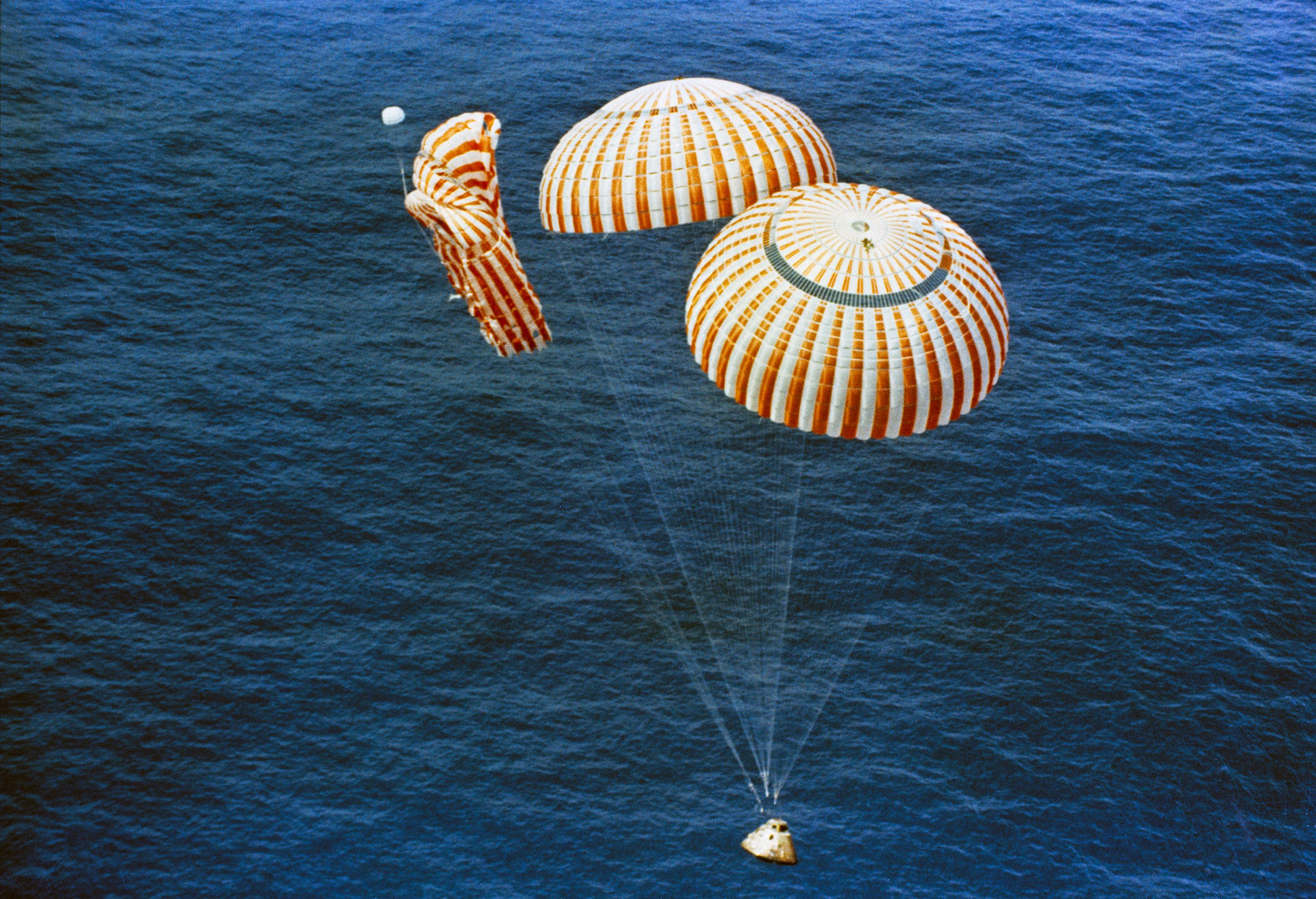}
    \caption{Die Kapsel der Apollo 15-Mission wassert mit nur zwei intakten Fallschirmen (Bild: NASA)}
    \label{fig:apollo}
\end{figure}

Landungen auf dem Mars gestalten sich jedoch anders. Seit Viking~1 werden die Sonden zwar durch Fallschirme abgebremst. Auf den letzten Kilometern wird die weiche Landung jedoch meistens durch Düsen bewerkstelligt. Auf diese Weise konnten das mobile Labor Curiosity und die geologische Messeinheit InSight erfolgreich auf dem Mars aufsetzen.

\begin{figure}[!ht]
    \centering
    \includegraphics[width=7cm]{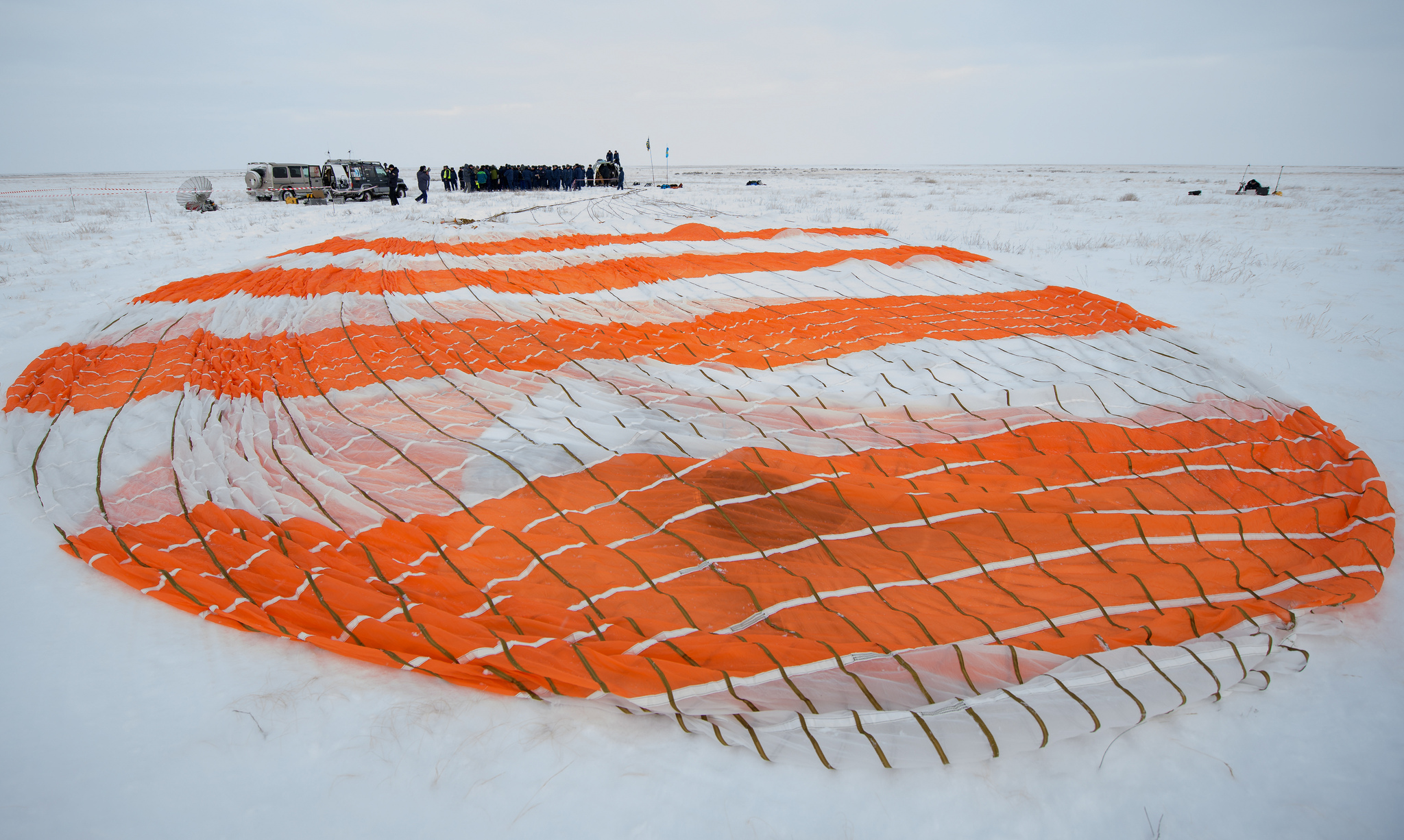}
    \caption{Der Fallschirm nach der Landung des Sojus MS-09-Raumschiffs, mit dem Alexander Gerst am 20. Dezember 2018 auf die Erde zurückkehrte (Bild: NASA, \url{https://www.flickr.com/photos/nasahqphoto/45477952845}, \href{https://creativecommons.org/licenses/by-nc-nd/2.0/legalcode}{CC BY-NC-ND 2.0}).}
    \label{fig:sojus}
\end{figure}

\begin{figure}[!ht]
    \centering
    \includegraphics[width=11cm]{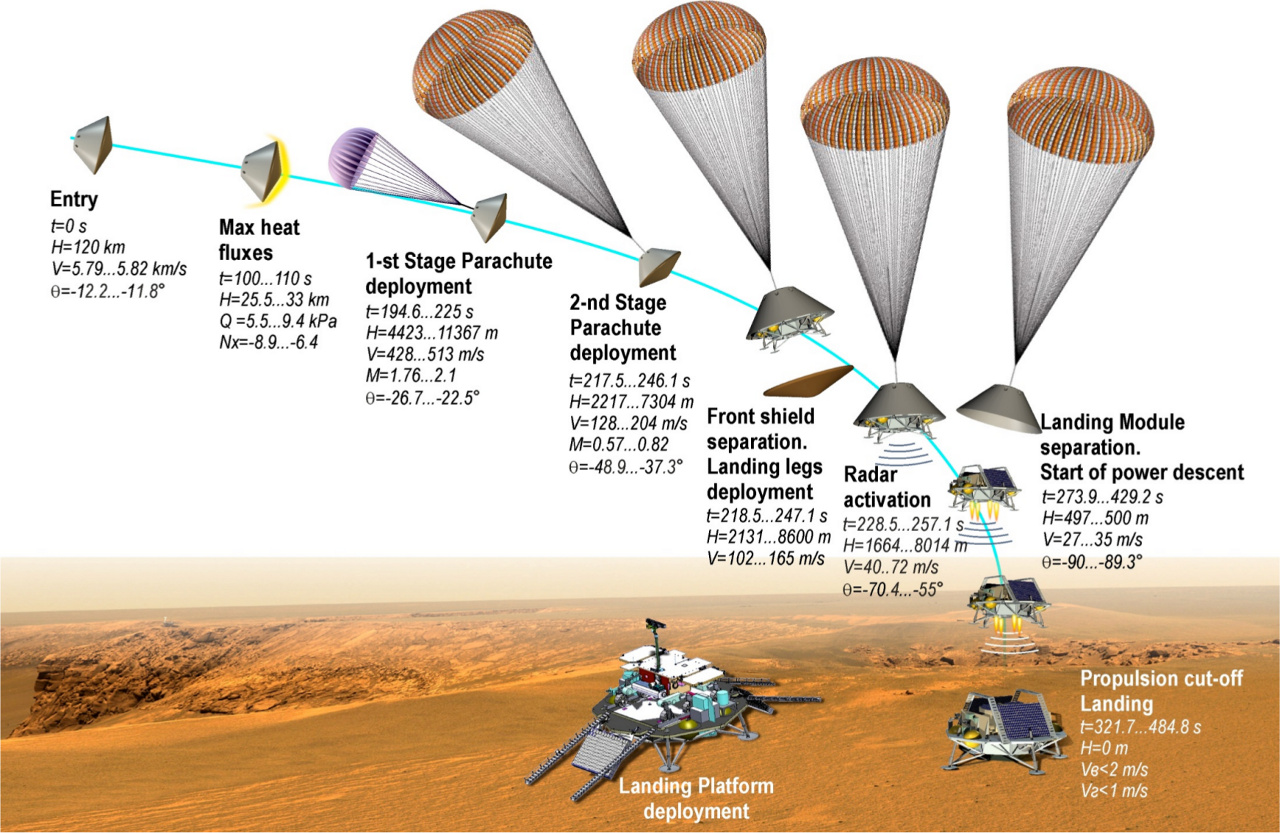}
    \caption{Schematische Darstellung der Landesquenz des ExoMars-Rovers. Mehrere große Fallschirme bremsen die Landeeinheit ab. Jedoch reicht die Reduktion der Geschwindigkeit für eine weiche Landung nicht aus. Die Sonde würde zerstört. Daher werden in der letzten Phase Triebwerke eingesetzt, so dass die Landeeinheit schließlich sanft aufsetzt (Bild: ESA).}
    \label{fig:sequenz}
\end{figure}

Auch der gemeinsam von ESA\footnote{European Space Agency, Europa} und Roskosmos\footnote{Russland} entwickelte ExoMars-Rover wird 2021 mit einer Landeeinheit durch Bremsraketen auf dem Mars landen (siehe Abb.~\ref{fig:sequenz}). Bei zukünftigen astronautischen Missionen wird es nicht anders sein.

\begin{figure}[!ht]
    \centering
    \includegraphics[width=11cm]{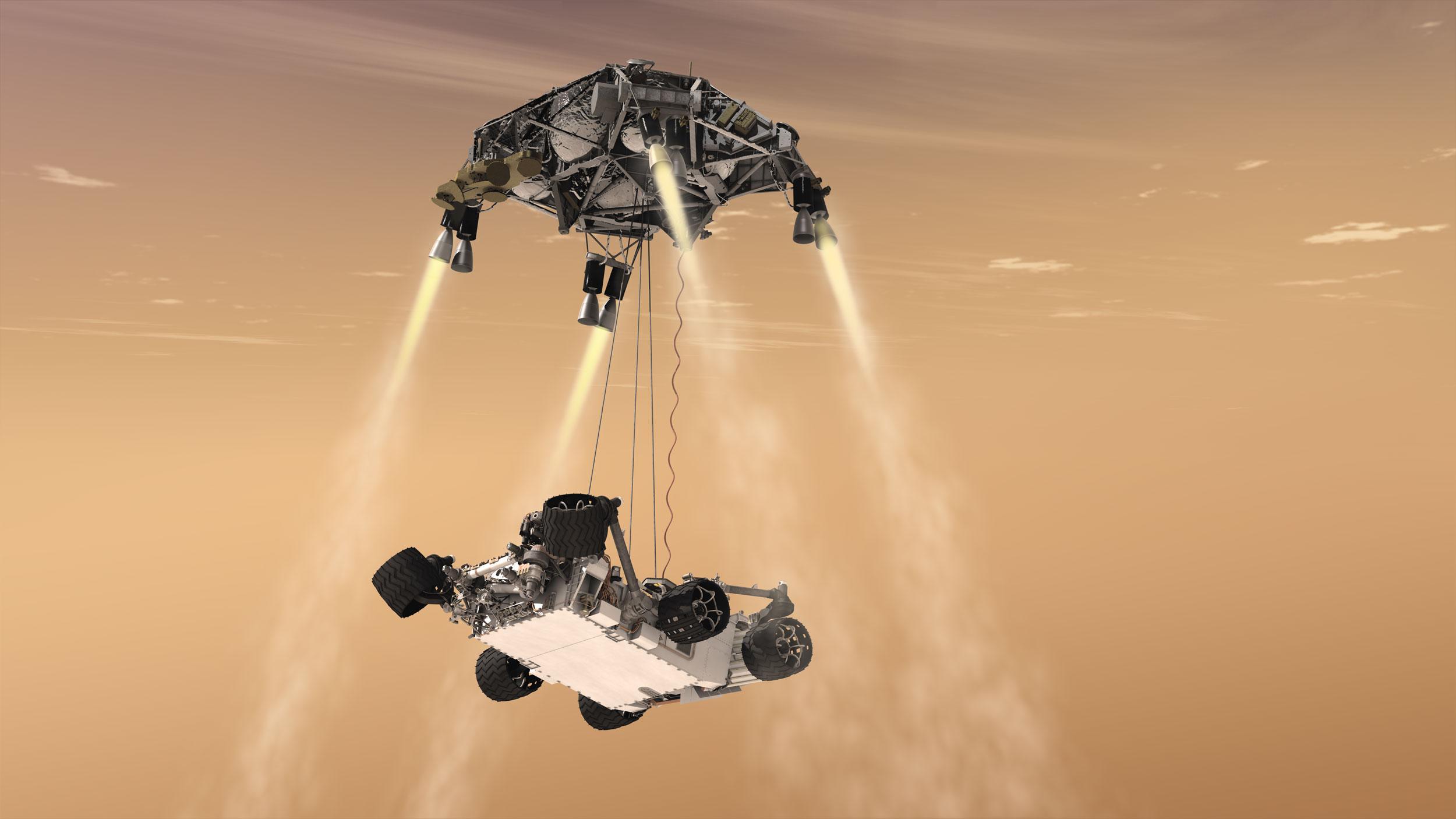}
    \caption{Künstlerische Darstellung der letzten Sequenz der Landung des Mars-Roboters „Curiosity“. Das Bremsmanöver wird nicht mit Fallschirmen sondern mit Triebwerken durchgeführt (Bild: NASA/JPL-Caltech).}
    \label{fig:curiosity}
\end{figure}

Doch warum ist das so? Die einfache Antwort ist: Die Atmosphäre ist zu dünn. Fallschirme alleine können die Sinkgeschwindigkeit nicht ausreichend reduzieren. Somit ist der Luftwiderstand zu gering, um genügend Reibung zu erzeugen. Die vom Fallschirm erzeugte Bremskraft reicht nicht aus, um die von der Gravitation ausgeübte Beschleunigung ausreichend zu kompensieren. Die Details werden im nächsten Abschnitt erläutert.

\subsection*{Physik des Fallschirms}
Der Fall mit Fallschirm ist eine Modifikation des freien Falls mit einer zusätzlichen zeitabhängigen Randbedingung durch ein bremsendes Medium. Dadurch entsteht eine Kraft, die der Beschleunigung entgegen wirkt. Im Vakuum existiert ein solches Medium jedoch nicht. Dann entspricht dies dem idealisierten freien Fall. Um die nachfolgenden Betrachtungen zu erleichtern, nehmen wir an, dass der Fall stets senkrecht zur Oberfläche, d. h. parallel zur Gravitationswirkung verläuft.

\subsubsection*{Freier Fall ohne Luftwiderstand}
Gemäß des Newtonschen Gravitationsgesetzes übt die Gravitationskraft $F_g$ auf einen Körper der Masse $m$ eine Beschleunigung $a$ aus. Daher kann man schreiben:

\begin{align*}
    F_g &= m\cdot a \\[5pt]
    \Leftrightarrow G\cdot \frac{M\cdot m}{r^2} &= m\cdot a\\[5pt]
    \Leftrightarrow a &=G\cdot \frac{M}{r^2} \equiv g
\end{align*}

\medskip
Dieser Term stellt eine allgemeine Bewegungsgleichung für eine gleichmäßig beschleunigte Bewegung dar. Hierbei ist $r$ der Abstand zwischen den Massenmittelpunkten. Daraus folgen per Integration die bekannten Gleichungen für die Geschwindigkeit und den Ort der beschleunigten Masse zum Zeitpunkt $t$.

\begin{equation}
v(t) = v_0 + G\cdot \frac{M}{r^2}\cdot t = v_0 + g\cdot t
\end{equation}

\begin{equation}
s(t) = s_0 + v_0\cdot t + \frac{1}{2}\cdot G\cdot \frac{M}{r^2}\cdot t^2 = s_0 + v_0\cdot t + \frac{1}{2}\cdot g\cdot t^2
\end{equation}

\medskip
Dabei stellen $s_0$ den Ort und $v_0$ die Geschwindigkeit zu Beginn der Beschleunigung dar.

\subsubsection*{Freier Fall mit Luftwiderstand}
\label{s:FFmL}
Die Kraft des Luftwiderstands\footnote{Herleitung unter: \url{https://www.leifiphysik.de/mechanik/reibung-und-fortbewegung/luftreibung}} wirkt in umgekehrter Richtung der Gravitationswirkung und beträgt:

\begin{equation}
    F_w = c_w \cdot \frac{1}{2} \cdot v^2 \cdot \varrho \cdot A
\end{equation}

\medskip
Diese Kraft\footnote{Bei laminarer Strömung folgt der Luftwiderstand aus dem Stokesschen Gesetz.} wird bestimmt durch Größen, die das Medium und das Objekt im Luftstrom charakterisieren. Auf der einen Seite ist die Kraft proportional zur Dichte des Mediums, auf der anderen Seite zur Querschnittsfläche, auf die die Kraft wirkt, sowie zum Quadrat der Strömungsgeschwindigkeit. Der Strömungswiderstandskoeffizient bzw. Widerstandsbeiwert $c_w$ ist eine dimensionslose Proportionalitätskonstante, die den Einfluss der Form des Objekts auf die Strömung und damit auf den Luftwiderstand beschreibt. Umgangssprachlich beziffert sie die Windschlüpfrigkeit. Sie kann durch kontrollierte Experimente z.~B. im Windkanal experimentell ermittelt werden.

\medskip
Die Dichte hängt zudem vom atmosphärischen Druck $p$ und der Temperatur $T$ ab. $R_S$ ist die spezifische Gaskonstante.

\begin{equation}
    \varrho = \frac{p}{R_S \cdot T}
\end{equation}

\medskip
Die Newtonsche Kraftgleichung lautet demnach:

\begin{align}
    m\cdot a &= F_g - F_w \\[5pt]
    \Leftrightarrow m\cdot a &= m\cdot g - c_w \cdot \frac{1}{2} \cdot v^2 \cdot \varrho \cdot A
\end{align}

\medskip
Aus der Erfahrung ist bekannt, dass der Fallschirmspringer nach dem Öffnen des Fallschirms eine konstante Endgeschwindigkeit $v_E$ erreicht -- im Gegensatz zur Situation des freien Falls im Vakuum. Wenn $v_E=\mathsf{const.}$, folgt $a=0$. Damit erhalten wir die Bestimmungsgleichung der Endgeschwindigkeit eines Fallschirmsprungs.

\begin{align}
  m\cdot a = 0 &= m\cdot g - c_w \cdot \frac{1}{2} \cdot v^2 \cdot \varrho \cdot A\\[5pt]
  \Leftrightarrow m\cdot g &= c_w \cdot \frac{1}{2} \cdot v^2 \cdot \varrho \cdot A\\[5pt]
  \Leftrightarrow v_E &= \sqrt{\frac{2\cdot m\cdot g}{c_w\cdot \varrho\cdot A}}\label{e:vE}
\end{align}

\medskip
Anders als beim freien Fall im Vakuum ist die Geschwindigkeit nicht unabhängig von der Masse des fallenden Gegenstands.

\medskip
Aus der Bewegungsgleichung folgen wiederum die Geschwindigkeits- und Ortsfunktionen. Setzen wir als Anfangsbedingungen $v_0=0$ und $s_0=0$, wird die Lösung übersichtlicher, und es folgt\footnote{Das zu lösende Integral enthält eine Funktion der Form $\frac{1}{1-x^2}$, deren Stammfunktion artanh ist.}:

\begin{equation}
    v(t) = v_E \cdot \tanh\left(\frac{g\cdot t}{v_E}\right)
\end{equation}

\begin{equation}
    s(t) = \frac{v_E^2}{g}\cdot\ln\left(\cosh\left(\frac{g\cdot t}{v_e}\right)\right)
\end{equation}

\medskip
Auch hier erkennt man:

\begin{equation}
    \lim_{t\rightarrow\infty} v(t) = v_E
\end{equation}

\subsection*{Fallschirme für Marsmissionen}
\label{s:fallschirme}
Um Sonden in der Marsatmosphäre abzubremsen, werden typischerweise mehrere verschiedene Fallschirme benutzt. So setzt man auch für die Landung des Rovers der ExoMars-Mission Schirme für die unterschiedlichen Geschwindigkeitsbereiche ein \autocite[][siehe auch Abb.~\ref{fig:sequenz}]{esa_exomars_2018}.

\medskip
Beim Eintritt in die Atmosphäre bremst das Landemodul zunächst durch die Reibung. In der nächsten Phase werden spezielle Überschall-Fallschirme entfaltet. Sie haben bei diesen Geschwindigkeiten Widerstandsbeiwerte von etwa $c_w=0,4$ \autocite[][]{takayanagi_development_2016}.

\medskip
In der letzten Phase bremsen besonders große Fallschirme die Landemodule auf geringe Geschwindigkeiten ab. Im Fall des ExoMars-Landemoduls kommt ein Fallschirm mit einem Durchmesser von \unit[35]{m} zum Einsatz \autocite[][]{esa_first_2018}. Ein typischer Widerstandsbeiwert für solche Fallschirme ist $c_w=1,28$ \autocite[][S.~47]{hillje_entry_1967}.

\clearpage
\section*{Vorbereitung}
Zur Einstimmung sollten sich die Schülerinnen und Schüler die nachfolgenden Videos ansehen. Sie zeigen Animationen von Landungen zweier Marssonden.

\bigskip
Curiosity has landed (Dauer: 2:30)\\
\url{https://youtu.be/N9hXqzkH7YA}

\bigskip
InSight: Landing on Mars (Dauer: 3:18)\\
\url{https://youtu.be/C0lwFLPiZEE}

\bigskip
Falls Sie die Eigenschaften des Mars nicht selber einführen, lassen Sie die Schülerinnen und Schüler die grundlegenden Charakteristika (z.~B.~Größe, Masse, Entfernung, Atmosphäre) selbst recherchieren. Hierzu bieten sich Artikel aus Wikipedia an. Nach dieser Einleitung lassen Sie die Schülerinnen und Schüler folgende Fragen beantworten. Ermuntern Sie sie zu Diskussionen.

\begin{itemize}
    \item Warum hilft ein Fallschirm, einen Absprung aus einem Flugzeug unbeschadet zu überleben?
    \item Wie entsteht die Bremswirkung eines Fallschirms?
    \item Warum funktioniert ein Fallschirm auf dem Mond nicht?
    \item Was muss man beim Einsatz eines Fallschirms auf dem Mars beachten, damit er gut funktioniert?
\end{itemize}

\section*{Aufgaben}
Die Schülerinnen und Schüler berechnen die Endgeschwindigkeiten für eine Landung auf dem Mars bzw. auf der Erde. Aus dem Vergleich wird ersichtlich, dass diese auf dem Mars für eine weiche Landung zu hoch ist. Bei den Berechnungen werden folgende vereinfachende Annahmen gemacht.

\begin{itemize}
\item Senkrechter Fall
\item Auftrieb vernachlässigt
\item Der Fallschirm hat keine Masse.
\item Konstante Gravitationsbeschleunigung während des Falls
\item Homogene Dichte der Atmosphäre während des Falls
\item Konstante atmosphärische Temperatur
\item Reine CO$_2$-Atmosphäre auf dem Mars
\item Ideales Gas
\end{itemize}

\newpage
\begin{table}[!ht]
    \centering
    \renewcommand{\arraystretch}{1.2}
    \caption{Wichtige Größen zur Berechnung der Aufgaben.}
    \label{tab:values}
    \begin{tabular}{lcc}
    \hline
    \multicolumn{1}{c}{Größe} & Formelzeichen & Wert \\
    \hline
Masse der ExoMars\footnote-Landeeinheit & $m$ & \unit[1140]{kg}\\
Schwerebeschleunigung Mars & $g_M$ & \unit[3,71]{$\frac{\mathsf{m}}{\mathsf{s}^2}$}\\
Schwerebeschleunigung Erde & $g_E$ & \unit[9,81]{$\frac{\mathsf{m}}{\mathsf{s}^2}$}\\
Widerstandsbeiwert des Fallschirms & $c_w$ & 1,28 \\
Querschnittsfläche des Fallschirms & $A$ & \unit[960]{$\mathsf{m}^2$}\\
Atmosphärischer Druck Mars & $p_M$ & \unit[636]{Pa}\\
Atmosphärischer Druck Erde & $p_E$ & \unit[101325]{Pa}\\
Dichte Luft bei Normaldruck & $\varrho_L$ & \unit[1,3]{$\frac{\mathsf{kg}}{\mathsf{m}^3}$}\\
Mittlere Temperatur Mars & $T_M$ & \unit[280]{K} \\
Spezielle Gaskonstante CO$_2$ & $R_{S,\mathsf{CO}_2}$ & \unit[188,9]{$\frac{\mathsf{J}}{\mathsf{kg}\cdot\mathsf{K}}$}\\
    \hline
    \end{tabular}
\end{table}

\footnotetext{ExoMars ist ein Marserkundungsprogramm von ESA (Europa) und Roskosmos (Russland).}

\subsection*{Endgeschwindigkeit eines Fallschirmsprungs}
Beim freien Fall im Vakuum beschleunigt ein Objekt gleichmäßig. Bei einem Fall mit einem Fallschirm innerhalb einer Atmosphäre strebt die Fallgeschwindigkeit jedoch einem konstanten Wert zu.

\medskip
Leite aus dem Prinzip

\begin{displaymath}
F = m\cdot a
\end{displaymath}

\medskip
die Gleichung für die Berechnung der Endgeschwindigkeit eines Falls am Fallschirm $v_E$ her.

\begin{displaymath}
v_E = \sqrt{\frac{2\cdot m\cdot g}{c_w\cdot \varrho\cdot A}}
\end{displaymath}

\medskip
Die wirkende Kraft $F$ setzt sich aus der Gravitationskraft
\begin{displaymath}
F_g=m\cdot g
\end{displaymath}

\medskip
und der Kraft des Luftwiderstands

\begin{displaymath}
F_w=c_w\cdot \frac{1}{2}\cdot v^2\cdot\varrho\cdot A
\end{displaymath}

\medskip
zusammen. Beide Kräfte wirken in entgegengesetzte Richtungen.

\subsection*{Landung am Fallschirm auf dem Mars}
Nutze die Daten aus Tab.~\ref{tab:values}, um die Endgeschwindigkeit des ExoMars-Rovers bei der Landung zu berechnen, die lediglich durch das Bremsen mit einem Fallschirm der Querschnittsfläche $A$ erzielt wird. Drücke das Ergebnis in den Einheiten m/s und km/h aus.

\medskip
Nimm zur Berechnung der Dichte der Marsatmosphäre an, dass diese ausschließlich aus Kohlenstoffdioxid (CO$_2$) besteht und die in der Tabelle angegebenen Eigenschaften hat. Der Zusammenhang zwischen Dichte und atmosphärischem Druck folgt aus der allgemeinen Gasgleichung für ideale Gase.

\begin{displaymath}
\varrho = \frac{p}{R_S \cdot T}
\end{displaymath}

\subsection*{Landung am Fallschirm auf der Erde}
Ermittle die Endgeschwindigkeit für ein Bremsmanöver mit Fallschirm auf der Erde.

\subsection*{Diskussion}
Wie in Abb.~\ref{fig:sequenz} gezeigt und auf Seite~\pageref{s:fallschirme} beschrieben, werden Sonden, die auf dem Mars landen, nacheinander mit verschiedenen Fallschirmen abgebremst. Erkläre, warum es für die Berechnung der Endgeschwindigkeit genügt, lediglich den letzten Fallschirm in dieser Sequenz zu betrachten.

\medskip
Erläutere den Unterschied der beiden berechneten Fälle. Beziehe dich hierbei auf die Eigenschaften der beiden Planeten und ihren Atmosphären.

\medskip
Was ist für die unterschiedlichen Geschwindigkeiten ausschlaggebend?

\medskip
Würde der ExoMars-Rover die Landung auf diese Weise überstehen? Überlege, was mit einem Auto geschähe, das mit derselben Geschwindigkeit gegen eine Wand fährt.

\medskip
Auf welche Weise verringert man die Sinkgeschwindigkeit, um eine weiche Landung auf dem Mars zu gewährleisten?

\subsection*{Dimension eines idealen Fallschirms}
Auf welche Querschnittsfläche bzw. welchen Durchmesser müsste man den Fallschirm auf dem Mars verändern, damit die Endgeschwindigkeit so groß wie auf der Erde ist? Ist das realistisch?

\clearpage
\section*{Lösungen}
\subsection*{Endgeschwindigkeit eines Fallschirmsprungs}
Die Herleitung an Stelle der Gl.~\ref{e:vE} auf Seite~\pageref{s:FFmL} erläutert.

\subsection*{Landung am Fallschirm auf dem Mars}
Man kann entweder die Funktion der Dichte $\varrho(p,T)$ in die Bestimmungsgleichung von $v_E$ einsetzen oder die Dichte auf dem Mars ausrechnen.

\begin{displaymath}
\varrho_M = \frac{p_M}{R_S \cdot T_M}=\frac{\unit[636]{Pa}}{\unit[188,9]{\frac{\mathsf{J}}{\mathsf{kg}\cdot\mathsf{K}}}\cdot \unit[280]{K}}=\unit[0,012]{\frac{\mathsf{kg}}{\mathsf{m}^3}}
\end{displaymath}

\begin{align*}
v_{E,M} &= \sqrt{\frac{2\cdot m\cdot g_M}{c_w\cdot \varrho_M\cdot A}}=\sqrt{\frac{2\cdot m\cdot g_M\cdot R_{S,\mathsf{CO}_2}\cdot T_M}{c_w\cdot p_M\cdot A}}\\[5pt]
&= \sqrt{\frac{2\cdot \unit[1140]{kg}\cdot\unit[3,71]{\frac{\mathsf{m}}{\mathsf{s}^2}}\cdot\unit[188,9]{\frac{\mathsf{J}}{\mathsf{kg}\cdot\mathsf{K}}}\cdot\unit[280]{K}}{1,28\cdot \unit[636]{Pa}\cdot\unit[960]{m^2}}}\\[5pt]
&= \unit[23,93]{\frac{m}{s}} = \unit[86,1]{\frac{km}{h}}
\end{align*}

\subsection*{Landung am Fallschirm auf der Erde}
\begin{align*}
v_{E,E} &= \sqrt{\frac{2\cdot m\cdot g_E}{c_w\cdot \varrho_L\cdot A}}=
\sqrt{\frac{2\cdot \unit[1140]{kg}\cdot\unit[9,81]{\frac{\mathsf{m}}{\mathsf{s}^2}}}{1,28\cdot\unit[1,3]{\frac{\mathsf{kg}}{\mathsf{m}^3}} \cdot\unit[960]{m^2}}}\\[5pt]
&= \unit[3,74]{\frac{m}{s}} = \unit[13,5]{\frac{km}{h}}
\end{align*}

\subsection*{Diskussion}
Die Endgeschwindigkeit hängt nicht von der Anfangsgeschwindigkeit ab, vorausgesetzt der Bremsvorgang dauert lange genug. Somit sind die Geschwindigkeiten, die durch die vorgeschalteten Fallschirme erzielt werden, für die Berechnung irrelevant.

\medskip
Die Endgeschwindigkeit ist auf dem Mars mehr als 6 mal so hoch wie auf der Erde. Obwohl die gravitative Anziehung auf dem Mars erheblich geringer als auf der Erde ist, fällt eine Sonde wie der ExoMars-Rover dort schneller zu Boden.

\medskip
Letztendlich ist die Atmosphäre auf dem Mars zu dünn, als dass sie einen ausreichend großen Widerstand auf den Fallschirm ausüben könnte. Die Erdatmosphäre ist bedeutend dichter.

\medskip
Der ExoMars-Rover würde den Aufprall mit einer Geschwindigkeit von \unit[86]{km/h} sicher nicht überstehen.

\medskip
Aus diesem Grund benötigt man auf dem Mars Triebwerke, um die Geschwindigkeit bei der Landung ausreichend zu verringern.

\subsection*{Dimension eines idealen Fallschirms}
\begin{displaymath}
v_E = \sqrt{\frac{2\cdot m\cdot g}{c_w\cdot \varrho\cdot A}} \Leftrightarrow
A = \frac{2\cdot m\cdot g}{c_w\cdot \varrho\cdot v_E^2}
\end{displaymath}

\medskip
Um nun die Querschnittsfläche des Fallschirms zu berechnen, setzen wir die Eigenschaften des Mars ein und nehmen als Endgeschwindigkeit $v_{E,E}$.

\begin{align*}
A_{id} &= \frac{2\cdot m\cdot g_M}{c_w\cdot \varrho_M}\cdot\frac{1}{v_{E.E}^2} = 
\frac{2\cdot m\cdot g_M}{c_w\cdot \varrho_M}\cdot\frac{c_w\cdot \varrho_L\cdot A}{2\cdot m\cdot g_E} =
\frac{\varrho_L\cdot g_M}{\varrho_M\cdot g_E}\cdot A\\[5pt]
&=\frac{\unit[1,3]{\frac{kg}{m^3}} \cdot \unit[3,71]{\frac{m}{s^2}}}{\unit[0,012]{\frac{kg}{m^3}} \cdot \unit[9,81]{\frac{m}{s^2}}}\cdot A = 40,89\cdot A = \unit[39251]{m^2}
\end{align*}

\medskip
Der Durchmesser eines runden Fallschirms mit dieser Querschnittsfläche beträgt:

\begin{displaymath}
d=2\cdot\sqrt{\frac{A}{\pi}}=\unit[224]{m}
\end{displaymath}

\medskip
Das ist völlig unrealistisch. Nicht nur kann man solch große Fallschirme nicht bauen und testen. Auch besitzen sie eine große Masse, die das Gesamtgewicht des Landemoduls enorm anwachsen ließe.

%%%%%%%%%%%%%%%%%%%%%%%%%%%%%%%%%%%%%%%%%%%%%%%%%%%%%%%%

\clearpage
% Danksagung
\section*{Danksagung}
Der Autor bedankt sich bei den Lehrern Matthias Penselin, Florian Seitz und Martin Wetz für ihre wertvollen Hinweise, Kommentare und Änderungsvorschläge, die in die Erstellung dieses Materials eingeflossen sind. Weiterer Dank gilt Herrn Dr. Volker Kratzenberg-Annies für seine gewissenhafte Durchsicht.

%BibLatex
\printbibliography

\clearpage
\ 
\vfill
\medskip
Diese Unterrichtsmaterialien sind im Rahmen des
Projekts {\em Raum für Bildung} am Haus der Astronomie in Heidelberg entstanden. Weitere Materialien des Projekts finden Sie unter:

\begin{center}
\href{http://www.haus-der-astronomie.de/raum-fuer-bildung}{http://www.haus-der-astronomie.de/raum-fuer-bildung}
und
\href{http://www.dlr.de/next}{http://www.dlr.de/next}
\end{center}

Das Projekt findet in Kooperation mit dem Deutschen Zentrum für Luft- und Raumfahrt statt und wird von der Joachim Herz Stiftung gefördert.\\
\begin{center}
\includegraphics[height=1.5cm]{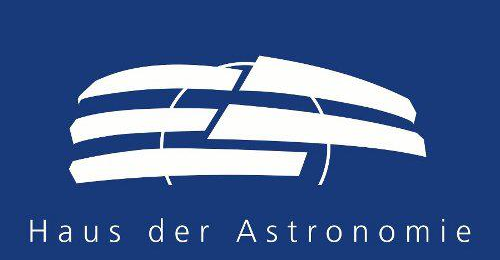}
\hspace*{4em}
\includegraphics[height=1.5cm]{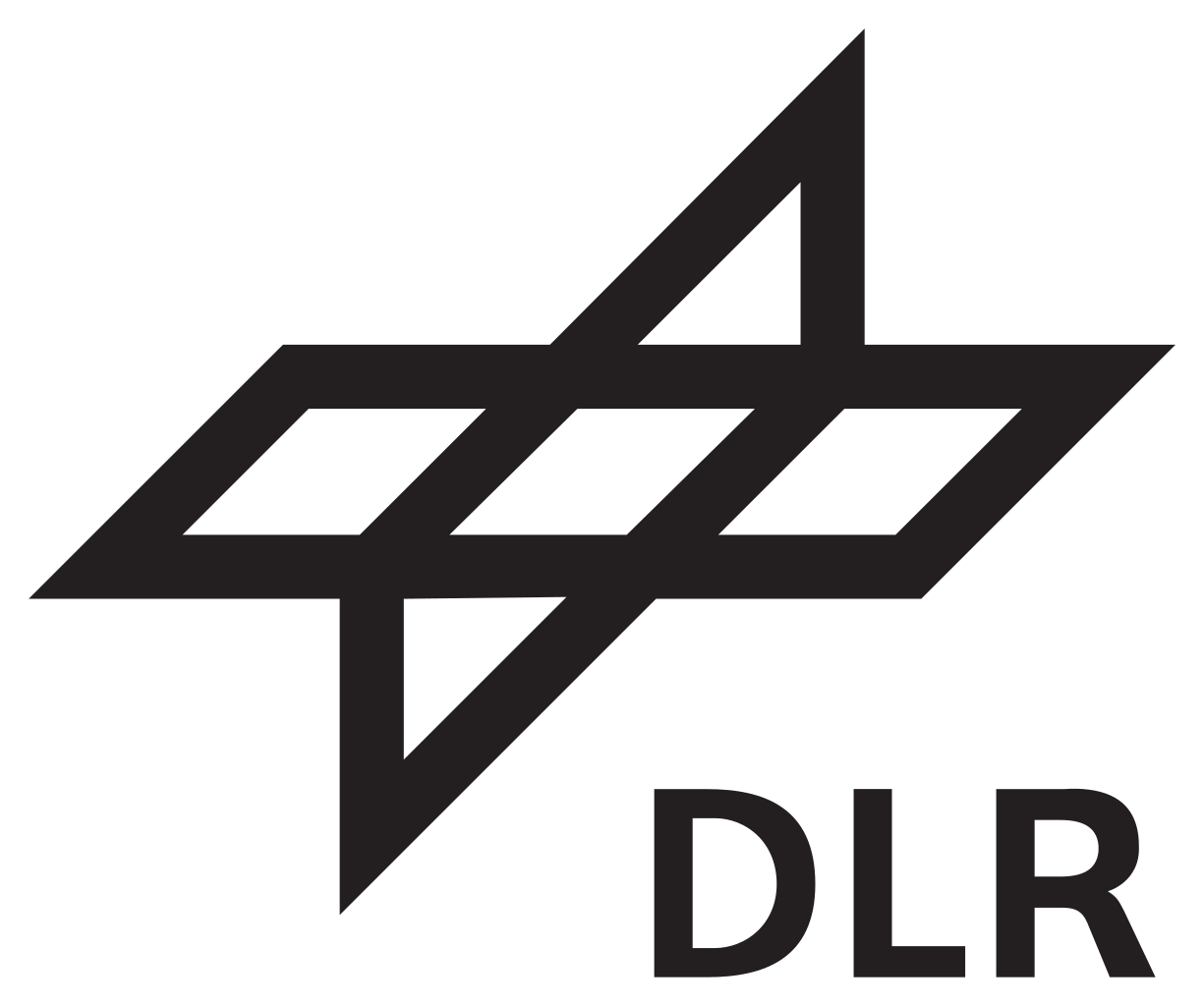}
\hspace*{4em}
\includegraphics[height=1.5cm]{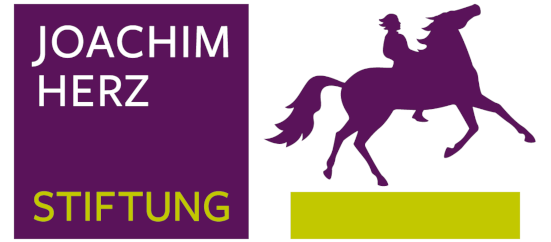}
\end{center}

\end{document}

%% file: header.tex
\usepackage[ngerman]{babel}
\usepackage[utf8]{inputenc}
\usepackage[T1]{fontenc}

\usepackage{amsmath}
\usepackage[slantedGreek]{cmbright}
\usepackage{units}
\usepackage{ellipsis}

\usepackage{tabularx}
\newcolumntype{L}[1]{>{\raggedright\arraybackslash}p{#1}}
\newcolumntype{C}[1]{>{\centering\arraybackslash}p{#1}}
\newcolumntype{R}[1]{>{\raggedleft\arraybackslash}p{#1}}

\usepackage[pdftex]{graphicx}
\usepackage[pdftex,colorlinks=true, allcolors=blue]{hyperref}
\usepackage[european]{circuitikz}
\usepackage[headsepline,footsepline]{scrlayer-scrpage}
\clearscrheadfoot

\usepackage[bf]{caption}
\newcommand\leftmarkline[1]{%
  \parbox[c][\layerheight][b]{\layerwidth}{%
    \hspace*{3.5mm}\rule{#1}{.2mm}%
}}
\DeclareNewLayer[{
  background,
  innermargin,
  oddpage,% bei zweiseitigen Dokumenten nur auf den ungeraden Seiten
  height=.34\paperheight,
  contents={\leftmarkline{2mm}}
}]{oberefaltmarke}
\DeclareNewLayer[{
  clone=oberefaltmarke,
  height=.67\paperheight
}]{unterefaltmarke}
\DeclareNewLayer[{
  clone=oberefaltmarke,
  height=.5\paperheight,
  contents={\leftmarkline{4mm}}
}]{lochermarke}
\AddLayersToPageStyle{@everystyle@}{lochermarke}

\setkomafont{pagefoot}{\normalfont\footnotesize}
\setkomafont{caption}{\normalfont\footnotesize}

\usepackage[babel,german=quotes]{csquotes} %deutsches Anführungszeichen

\usepackage[backend=biber,style=authoryear-icomp]{biblatex}
\newcommand{\myproject}{\Large Unterrichtsmaterialien zur\\ ISS-Horizons-Mission von Alexander Gerst}
\newcommand{\mytitle}{Weiche Landung auf dem Mars}
\newcommand{\myclasses}{Klassen 9-13}
\newcommand{\myauthor}{Markus Nielbock}

\rehead*{\small In Kooperation mit\\ \includegraphics[height=11mm]{DLR_Logo_svg.png}}
\rohead*{\small In Kooperation mit\\ \includegraphics[height=11mm]{DLR_Logo_svg.png}}
\lehead*{\includegraphics[height=15mm]{HdA_500x260.png}}
\lohead*{\includegraphics[height=15mm]{HdA_500x260.png}}
\ofoot*{Seite \pagemark}
\ifoot*{Handreichung: \mytitle}

%% file: titlepage.tex
%%%%%%%%%%%%%%%%%%%%%%%%% Titlepage %%%%%%%%%%%%%%%%%%%%%%%%%%%%%%%%%%%%
\begin{titlepage}
\thispagestyle{scrheadings}

\begin{center}
{\large\bfseries
Handreichung für Lehrpersonen:

\bigskip
\LARGE
\mytitle
}

\bigskip
{\large\bfseries
\myclasses
}

 \bigskip
{\bfseries
\myauthor
}

\bigskip
30.~Januar~2019
\end{center}

\section*{Zusammenfassung}
In ferner Zukunft werden sicher Menschen zum Mars fliegen. Schon heute führen Sonden und Fahrzeuge auf der Marsoberfläche Messungen durch. Landungen können auf dem Mars jedoch meistens nicht alleine mit Fallschirmen durchgeführt werden. Stattdessen verwendet man Landemodule, die mit Triebwerken die Geschwindigkeit auf akzeptable Werte verringern. Im Vergleich zur Erde übt der Mars zwar eine geringere Anziehungskraft auf die Sonden aus. Jedoch ist seine Atmosphäre sehr viel dünner als die der Erde. Mit simplen Rechnungen und einigen vereinfachenden Annahmen berechnen Schülerinnen und Schüler die Geschwindigkeit einer Marssonde mit Fallschirm. Dadurch erkennen sie, dass aufgrund der dünnen Marsatmosphäre die Endgeschwindigkeit keine weiche Landung zulässt.

\section*{Lernziele}
Die Schülerinnen und Schüler
\begin{itemize}
\item ermitteln eine Bestimmungsgleichung für die Endgeschwindigkeit eines Falls mit Fallschirm,
\item berechnen die Endgeschwindigkeiten von Fallschirmabwürfen auf dem Mars und auf der Erde,
\item erläutern die Gründe dafür, dass Fallschirme alleine keine weiche Landung zulassen,
\item berechnen die Größe eines Fallschirms, der eine weiche Marslandung ermöglichen würde.
\end{itemize}

\section*{Materialien}
\begin{itemize}
\item Arbeitsblätter (erhältlich unter \href{http://www.haus-der-astronomie.de/raum-fuer-bildung}{http://www.haus-der-astronomie.de/raum-fuer-bildung})
\item Stift
\item Taschenrechner
\item Computer/Tablet/Smartphone mit Internetzugang (optional)
\end{itemize}

\section*{Stichworte}
Mars, Erkundung, Reibung, Luftwiderstand, Luftdruck, Dichte, Strömung 

\section*{Dauer}
90 Minuten

% \vfill
% \begin{center}
% \centering
% \includegraphics[width=0.3\textwidth]{Horizons_logo.png}

% \footnotesize
% Bild: ESA/Steinbeis Beratungszentrum/Hochschule Darmstadt
% \end{center}

\end{titlepage}

%% file: mars.bib
@report{hillje_entry_1967,
	location = {Houston, Texas, {USA}},
	title = {{ENTRY} {FLIGHT} {AERODYNAMICS} {FROM} {APOLLO} {MISSION} {AS}-202},
	url = {https://ntrs.nasa.gov/archive/nasa/casi.ntrs.nasa.gov/19670027745.pdf},
	number = {{NASA} {TN} D-4185},
	institution = {{NASA} Manned Spacecraft Center},
	type = {{NASA} Technical Note},
	author = {Hillje, Ernest R.},
	date = {1967-10},
	langid = {american}
}

@online{esa_first_2018,
	title = {First test success for largest Mars mission parachute},
	url = {http://exploration.esa.int/mars/60118-first-test-success-for-largest-mars-mission-parachute/},
	titleaddon = {Robotic Exploration of Mars},
	author = {{ESA}},
	urldate = {2019-01-06},
	date = {2018-03-29},
	langid = {british}
}

@online{esa_exomars_2018,
	title = {{ExoMars} 2020 parachute deployment sequence},
	url = {http://exploration.esa.int/mars/60120-exomars-2020-parachute-deployment-sequence/},
	titleaddon = {Robotic Exploration of Mars},
	author = {{ESA}},
	urldate = {2019-01-06},
	date = {2018-03-29},
	langid = {british}
}

@article{takayanagi_development_2016,
	title = {Development of Supersonic Parachute for Japanese Mars Rover Mission},
	volume = {14},
	issn = {1884-0485},
	url = {https://www.jstage.jst.go.jp/article/tastj/14/ists30/14_Pe_87/_article/-char/en},
	doi = {10.2322/tastj.14.Pe_87},
	pages = {Pe\_87--Pe\_94},
	issue = {ists30},
	journaltitle = {{TRANSACTIONS} {OF} {THE} {JAPAN} {SOCIETY} {FOR} {AERONAUTICAL} {AND} {SPACE} {SCIENCES}, {AEROSPACE} {TECHNOLOGY} {JAPAN}},
	shortjournal = {{AEROSPACE} {TECHNOLOGY} {JAPAN}},
	author = {Takayanagi, Hiroki and Suzuki, Toshiyuki and Yamada, Kazuhiko and Maru, Yusuke and Matsuyama, Shingo and Fujita, Kazuhisa},
	urldate = {2019-01-06},
	date = {2016},
	langid = {english}
}

@mvbook{meyer_meyers_1885,
	location = {Leipzig},
	edition = {4},
	title = {Meyers Konversations-lexikon. Eine Encyklopädie des allgemeinen Wissens},
	volume = {11},
	rights = {http://creativecommons.org/publicdomain/mark/1.0/},
	url = {http://archive.org/details/bub_gb_IpEGAQAAIAAJ},
	volumes = {19},
	pagetotal = {1162},
	publisher = {Verlag des Bibliographischen Instituts},
	author = {Meyer, Herrmann Julius},
	editora = {{University of California}},
	editoratype = {collaborator},
	urldate = {2019-01-04},
	date = {1885},
	langid = {german}
}

@online{chilton_war_2016,
	title = {The War of the Worlds panic was a myth},
	url = {https://www.telegraph.co.uk/radio/what-to-listen-to/the-war-of-the-worlds-panic-was-a-myth/},
	titleaddon = {The Telegraph},
	author = {Chilton, Martin},
	urldate = {2019-01-14},
	date = {2016-05-06},
	langid = {british}
}

@article{schiaparelli_decouvertes_1882,
	title = {Découvertes Nouvelles sur la Planète Mars},
	volume = {1},
	url = {http://esoads.eso.org/cgi-bin/nph-data_query?bibcode=1882LAstr...1..216S&link_type=ARTICLE&db_key=AST&high=},
	pages = {216},
	journaltitle = {L'Astronomie},
	shortjournal = {{LAstr}},
	author = {Schiaparelli, Giovanni Virgilio},
	urldate = {2019-01-14},
	date = {1882-08},
	langid = {french}
}

@article{schiaparelli_observations_1897,
	title = {Observations de la Planète Mars},
	volume = {11},
	url = {http://esoads.eso.org/cgi-bin/nph-data_query?bibcode=1897BSAFR..11..107S&link_type=ARTICLE&db_key=AST&high=},
	pages = {107},
	journaltitle = {Bulletin de la Societe Astronomique de France et Revue Mensuelle d'Astronomie, de Meteorologie et de Physique du Globe},
	shortjournal = {{BSAFR}},
	author = {Schiaparelli, Giovanni Virgilio},
	urldate = {2019-01-14},
	date = {1897},
	langid = {french}
}

@online{zundel_mars-mission_2016,
	title = {Mars-Mission "{ExoMars}" – Suche nach Leben auf dem roten Planeten},
	url = {https://www.daserste.de/information/wissen-kultur/w-wie-wissen/weltall-218.html},
	titleaddon = {[W] wie Wissen - Das Erste},
	type = {Elektronisches Dokument},
	author = {Zündel, Wolfgang},
	urldate = {2019-01-14},
	date = {2016-12-10},
	langid = {german}
}
